\documentclass[twocolumn]{article}
\begin{document}

\title{Characterization of a noisy quantum process
\\
by complementary classical operations
}

\author{Holger F. Hofmann$^1$, Ryo Okamoto$^2$ and
Shigeki Takeuchi$^2$\\
$^1$Graduate School of Advanced Sciences of Matter, Hiroshima University\\ Kagamiyama 1-3-1, Higashi Hiroshima 739-8530, Japan
\\
$^2$Research Institute for Electronic Science,
Hokkaido University
\\
Sapporo 060-0812, Japan
}

\date{}

\maketitle

\abstract{
One of the challenges in quantum information is the
demonstration of quantum coherence in the operations of
experimental devices. While full quantum process tomography
can do the job, it is both cumbersome and unintuitive.
In this presentation, we show that a surprisingly detailed
and intuitively accessible characterization of errors is
possible by measuring the error statistics of only two
complementary classical operations of a quantum gate.
}


\section{Introduction}

Due to rapid technological advances, there is now an
increasing number of experimental realizations of
multi qubit quantum processes.
As more and more complex processes become feasible, it is
desirable to develop efficient strategies for the
characterization of such gates. Specifically, it may be
useful to identify the noise characteristics of a gate without
having to perform the large number of measurements necessary
for full quantum process tomography.

A particularly promising approach appears to be the analysis of
a pair of complementary classical operations performed by the
quantum device \cite{Hof05,Oka05,Hof06a,Hof06b}.
The errors observed in such a pair of classical operations provide
a surprisingly detailed ``map'' of the noise in the complete
quantum operation. It is therefore possible to use this limited
set of measurement results to estimate the performance of the
gate in operations that have not been evaluated explicitly.

In the following, we will briefly explain why the coherence of a
quantum operation is completely described by the "parallel"
performance of a pair of complementary operations, and how the
observable error syndromes correspond to the process matrix
elements describing the complete quantum process. We then apply
our analysis to the actual experimental data obtained from a
recently realized optical quantum controlled-NOT gate.
In particular, we can derive estimates for the success of
entanglement generation and of Bell state analysis, even though
the operations actually measured were completely local.

\section{Error classification}

In general, any noisy quantum process in a $d$ dimensional Hilbert
space can be described using a set of $d^2$ orthogonal operators
$\hat{\Lambda}_i$ by treating the operators as vectors in an
operator space with an inner product defined by the product trace
of two operators. If the intended operation is given by the
operator $\hat{U}_0$, errors may be classified according to deviations from the intended output by multiplying an orthogonal
set of output error operators $\hat{\Lambda}_i$ with the
ideal operation $\hat{U}_0$. A specific process $E$ is then
described by a process matrix with elements $\chi_{ij}$,
so that the input-output relation is given by
\begin{equation}
\label{eq:pmat}
\hat{\rho}_{\mbox{out}} = E(\hat{\rho}_{\mbox{in}})
= \sum_{i,j} \chi_{ij} \hat{\Lambda}_i \hat{U}_0\hat{\rho}_{\mbox{in}}
\hat{U}_0^\dagger \hat{\Lambda}_j.
\end{equation}

In principle, any choice of operator basis $\hat{\Lambda}_i$
is allowed. For practical purposes, however, $\hat{\Lambda}_i$
should be as close to the directly observed errors as possible.
For operations on quantum bits, the process should therefore
be expanded in terms of the error syndromes given by the Pauli
matrices $I$, $X$, $Y$, and $Z$, acting independently on each
qubit. Since quantum information processes are usually defined
in the computational basis given by the $Z$ eigenstates
(the $Z$ basis, for short), it is possible to identify $X$ with
a bit flip error, $Z$ with a phase error, and $Y=i XZ$ with
a combined bit flip and phase error \cite{Nie}. However, the
symmetry of the operations suggests that it may be more realistic
to interpret each operator as a Bloch vector rotation by $\pi$
around the appropriate axis. It is then clear that a phase error
in the $Z$ basis will show up as a bit flip in the $X$ basis,
ans vice versa. Since experiments on individual qubits can
only measure the bit value of a given basis, it is more realistic
to regard ``phase errors'' as observable bit flips in the
complementary basis. It is then possible to identify each error
operator by a pair of observable error patterns, one in the
computational $Z$ basis and another in the complementary $Z$ basis.

\begin{table}[ht]
\caption{Examples of error indices based on the error patterns observed
in the complementary $Z$ and $X$ basis.}
\label{tbl1}
\begin{center} 
\begin{tabular}{c|cc|c}
Error $\hat{\Lambda}_i$ & $Z$ error $f_z$& $X$ error $f_x$ &
index $i$
\\ \hline
$I \otimes X \otimes Y$ & $011 = 3$ & $001 = 1$ & $3, 1$
\\
$Z \otimes Y \otimes Y$ & $011 = 3$ & $111 = 7$ & $3, 7$
\\
$Y \otimes I \otimes Y$ & $101 = 5$ & $101 = 5$ & $5, 5$
\end{tabular}
\normalsize
\end{center}
\end{table}
A complete set of $d^2=2^{2N}$ orthogonal errors $\hat{\Lambda}_i$
is thus obtained by simply combining the $2^N$ possible $N$ qubit
errors in $Z$ with the $2^N$ possible errors in $X$.
For convenience, the index $i$ can then be written as $i=(f_z,f_x)$,
where $f_{z/x}$ is the value of the binary number obtained by
assigning a digit of $1$ to the locations of bit flip errors.
Table \ref{tbl1} shows some examples of this labeling system
for three qubit errors ($64$ possibilities).

\section{Evaluation of experimental results }

An experimental quantum gate can now be tested by using only
two complementary sets of inputs, $\{\mid n \rangle\}$ and
$\{\mid k \rangle\}$, chosen in such a way that the correct
outputs of the $d=2^N$ orthogonal states $\mid n \rangle$ are the
eigenstates of $Z$, $\{\hat{U}_0 \mid n \rangle =
\mid Z_n \rangle\}$, and the correct outputs of the $d=2^N$
orthogonal states $\mid k \rangle$ are the
eigenstates of $X$, $\{\hat{U}_0 \mid k \rangle =
\mid X_k \rangle\}$. The experimental results can be represented
in a pair of error tables showing the probabilities for the
various outcomes. Each outcome can then be identified by its
error number $f_{z/x}$, so that the correct output $Z_n$ ($X_k$) is given
by the conditional probability $p(0|Z_n)$ ($p(0|X_k)$), and
the observed error probabilities are given by
\begin{eqnarray}
\label{eq:pcond}
p(f_z|Z_n) &=& \langle Z_n \mid \hat{\Lambda}_{f_z,0}
E(\mid n \rangle \langle n \mid) \hat{\Lambda}_{f_z,0}
\mid Z_n \rangle
\nonumber \\
p(f_x|X_k) &=& \langle X_k \mid \hat{\Lambda}_{0,f_x}
E(\mid k \rangle \langle k \mid) \hat{\Lambda}_{0,f_x}
\mid X_k \rangle.
\nonumber
\\
\end{eqnarray}
Outcomes corresponding to the same kind of error can then
be averaged to obtain the
fidelities $F_Z$ and $F_X$ and the error probabilities
$\eta_Z(f_z)$ and $\eta_X(f_x)$ of the $Z$ and $X$ operations,
respectively. Specifically, the sums defining the fidelities and
the error probabilities read
\begin{eqnarray}
\label{eq:eval}
F_Z = \frac{1}{d} \sum_{Z_n=0}^{d-1} p(0|Z_n) &&
\eta_Z(f_z) = \frac{1}{d} \sum_{Z_n=0}^{d-1} p(f_z|Z_n)
\nonumber \\
F_X = \frac{1}{d} \sum_{X_k=0}^{d-1} p(0|X_k) &&
\eta_X(f_x) = \frac{1}{d} \sum_{X_k=0}^{d-1} p(f_x|X_k).
\nonumber \\
\end{eqnarray}
\begin{table*}[ht]
\caption{Measurement data for an experimental two qubit gate
\cite{Oka05} arranged according to the observed output errors.
The error distribution is summarized by the averages at the
bottom of each table.}
\label{tbl2}
\begin{center} \large
\begin{tabular}{l|cccc|}
$p(f_z|Z_n)$ & $f_z=0$ & $f_z=1$ & $f_z=2$ & $f_z=3$
\\ \hline
$Z_n=00$ & 0.898 & 0.031 & 0.061 & 0.011
\\
$Z_n=01$ & 0.885 & 0.021 & 0.088 & 0.006
\\
$Z_n=10$ & 0.819 & 0.054 & 0.031 & 0.096
\\
$Z_n=11$ & 0.810 & 0.099 & 0.027 & 0.064
\\ \hline &&&& \\[-0.2cm] averages
& $F_Z=0.853$ & $\eta_Z(1)=0.051$ &
$\eta_Z(2)=0.052$ & $\eta_Z(3)=0.044$
\\
\multicolumn{5}{c}{\vspace{0.5cm}}
\\
$p(f_x|X_n)$ & $f_x=0$ & $f_x=1$ & $f_x=2$ & $f_x=3$
\\ \hline
$X_k=00$ & 0.854 & 0.044 & 0.063 & 0.039
\\
$X_k=01$ & 0.870 & 0.019 & 0.071 & 0.040
\\
$X_k=10$ & 0.871 & 0.058 & 0.050 & 0.021
\\
$X_k=11$ & 0.874 & 0.013 & 0.099 & 0.013
\\ \hline &&&& \\[-0.2cm] averages
& $F_X=0.867$ & $\eta_X(1)=0.034$ &
$\eta_X(2)=0.071$ & $\eta_X(3)=0.028$
\end{tabular}
\normalsize
\end{center}
\end{table*}
Table \ref{tbl2} shows the experimental data reported for an
optical quantum controlled-NOT \cite{Oka05}
arranged according to the errors $f_z$ and $f_x$.
Note that the performance of the
experimental device is now described by only two fidelities
and $2 d-2$ error probabilities. For the quantum controlled-NOT
and other two qubit operations, this means that only $8$
characteristic probabilities are used to evaluate a process
fully described by a total of 256 process matrix elements.

Classical intuition already indicates that the probabilities
of the errors $f_z$ observed in $Z$ should correspond to the
sums of the diagonal process matrix elements with the same
value of $f_z$ in the first part of the index $i$, and the
probabilities observed in $X$ should correspond to the sums
diagonal process matrix elements with the same $f_x$.
This relation can indeed be confirmed by applying the
process matrix definition in eq.(\ref{eq:pmat}) to the
averages of the conditional probabilities defined by
eqs. (\ref{eq:pcond}) and (\ref{eq:eval}). As expected,
the results read
\begin{eqnarray}
\label{eq:relate}
F_Z = \sum_{f_x=0}^{d-1} \chi_{(0,f_x)(0,f_x)}, &&\hspace*{-0.5cm}
\eta_Z(f_z) = \sum_{f_x=0}^{d-1} \chi_{(f_z,f_x)(f_z,f_x)},
\nonumber \\
F_X = \sum_{f_z=0}^{d-1} \chi_{(f_z,0)(f_z,0)}, &&\hspace*{-0.5cm}
\eta_X(f_x) = \sum_{f_z=0}^{d-1} \chi_{(f_z,f_x)(f_z,f_x)}.
\nonumber \\
\end{eqnarray}
It is therefore possible to interpret the diagonal elements
$\chi_{(f_z,f_x)(f_z,f_x)}$ as joint probabilities of
the errors $f_x$ and $f_z$, where the unconditional
probabilities of $f_z$ are given by the measurement results
$F_Z$ and $\eta_Z(f_z)$, and the unconditional
probabilities of $f_x$ are given by the measurement results
$F_X$ and $\eta_X(f_x)$.
The process matrix elements can then be estimated by making
some additional assumptions about the correlations of
errors in $X$ and in $Z$.

The relation between the observed error distributions and
the diagonal elements of the density matrix can be visualized
by arranging the diagonal elements $\chi_{(f_z,f_x)(f_z,f_x)}$
in a table so that the lines represent the errors $f_z$
observed in $Z$ and the columns represent the errors $f_x$
observed in $X$. Table \ref{tbl3} shows this arrangement for
a two qubit operation such as the quantum controlled-NOT
that was first analyzed by this method in \cite{Oka05}. The experimentally
observed fidelities and error probabilities are then given by
the sums of the corresponding column or line.

Since all diagonal elements of the process matrix must be
positive, the measurement values impose rather strict limitations
on various properties of the gate. Of particular interest
the estimate of the process fidelity $\chi_{(0,0)(0,0)}=F_{qp}$,
since it provides a measure
of how close the experimental gate is to the intended ideal
process defined by $\hat{U}_0$. The relation between the process
fidelity $F_{qp}$ and the experimentally observed complementary
fidelities $F_Z$ and $F_X$ can be determined directly from
eq.(\ref{eq:relate}), making use of the fact that the trace of
the process matrix is one. The sum of the two complementary
fidelities then reads
\begin{equation}
\label{eq:Fsum}
F_Z+F_X = (F_{qp} + 1) - \left(\sum_{f_x=1}^{d-1}
\sum_{f_z=1}^{d-1}\chi_{(f_z,f_x)(f_z,f_x)}\right).
\end{equation}
Since the sum over process matrix elements with both $f_z \neq 0$
and $f_x \neq 0$ is always positive, the sum of $F_Z$ and $F_X$ can
never be larger than the process
fidelity plus one. At the same time, eq.(\ref{eq:relate}) ensures that
no classical fidelity is smaller than the process fidelity.
For this reason, the two measurement results $F_Z$ and $F_X$
limit the process fidelity of an experimental device to the
interval given by \cite{Hof05}
\begin{equation}
F_Z+F_X-1 \leq F_{qp} \leq \mbox{Min}\{F_Z,F_X\}.
\end{equation}
The difference between the upper and the lower limit of this
interval is equal to $\mbox{Max}\{F_Z,F_X\}-1$. For devices
with at least one very high fidelity, this estimate is therefore
nearly as good as a direct measurement of the process fidelity.

\begin{table*}[ht]
\caption{Illustration of the limits on diagonal process matrix
elements defined by the measurement results for the complementary
$Z$ and $X$ operations for a general 2 qubit operation.}
\label{tbl3}
\begin{center} \large
\begin{tabular}{l|cccc|c}
$\chi_{(f_z,f_x)(f_z,f_x)}$ & $f_x=0$ & $f_x=1$ & $f_x=2$ & $f_x=3$ & Sum
\\ \hline &&&& \\[-0.4cm]
$f_z=0$ & $F_{qp}$ & $\chi_{(0,1)(0,1)}$ & $\chi_{(0,2)(0,2)}$
& $\chi_{(0,3)(0,3)}$ & $F_Z$
\\
$f_z=1$ & $\chi_{(1,0)(1,0)}$ & $\chi_{(1,1)(1,1)}$ &
$\chi_{(1,2)(1,2)}$ & $\chi_{(1,3)(1,3)}$ & $\eta_Z(1)$
\\
$f_z=2$ & $\chi_{(2,0)(2,0)}$ & $\chi_{(2,1)(2,1)}$ &
$\chi_{(2,2)(2,2)}$ & $\chi_{(2,3)(2,3)}$ & $\eta_Z(2)$
\\
$f_z=3$ & $\chi_{(3,0)(3,0)}$ & $\chi_{(3,1)(3,1)}$ &
$\chi_{(3,2)(3,2)}$ & $\chi_{(3,3)(3,3)}$ & $\eta_Z(3)$
\\[0.1cm] \hline &&&& \\[-0.4cm] Sum
& $F_X$ & $\eta_X(1)$ &
$\eta_X(2)$ & $\eta_X(3)$ & 1
\end{tabular}
\normalsize
\end{center}
\end{table*}

In principle, the process fidelity can then be used to estimate
important gate properties such as the entanglement capability
of the gate \cite{Oka05}. However, more precise information
is available if the complete error statistics are used to
estimate the fidelities of operations other than the $Z$ and
$X$ operations observed in the experiment \cite{Hof06a,Hof06b}.
For this purpose, it is useful to construct a ``worst case''
error model, where all errors are either pure $Z$ errors
($i=(f_z,0)$) or pure $X$-errors ($i=(0,f_x)$).
The diagonal elements of the process matrix are then given
by the corresponding measurement results,
\begin{eqnarray}
\label{eq:lowF}
\chi_{(f_z,0)(f_z,0)} &=& \eta_Z(f_z),
\nonumber \\
\chi_{(0,f_x)(0,f_x)} &=& \eta_X(f_x),
\end{eqnarray}
and the process fidelity has its minimal value of
$F_{qp}=F_Z+F_X-1$. In the case of the quantum controlled-NOT
gate \cite{Oka05}, the process matrix elements defined by
this noise model are shown in table \ref{tbl4}.

\section{Noise models and fidelity estimates}

It is now possible to make predictions for other operations
based on this noise model.
In particular, any operation resulting in local output states
that are eigenstates of some combination of $X$, $Y$, and $Z$
eigenstates has a fidelity given by a well defined sum of
$d=2^N$ diagonal elements of the process matrix, corresponding
to the error operators that stabilize the output states.
For example, the two qubit operation resulting in $ZX$ outputs
has a fidelity of
\begin{equation}
\label{eq:fzx}
F_{zx} = F_{qp} + \chi_{(1,0)(1,0)} + \chi_{(0,2)(0,2)}
+ \chi_{(1,2)(1,2)}
\end{equation}
In the case of the quantum controlled-NOT, the $ZX$ operation
is simply the identity operation, since the $ZX$ eigenstates
are also eigenstates of the ideal controlled-NOT
operation $\hat{U}_0$. We can therefore estimate how well
the quantum gate preserves its eigenstates.
In terms of the error probabilities $\eta_{x/z}(F_{x/z})$,
the result reads
\begin{equation}
F_{zx} \geq 1 - \eta_Z(2) - \eta_Z(3) - \eta_X(1) -\eta_X(3) = 0.842.
\end{equation}
Note that this result is significantly larger than the result
of $0.72$ defined by the minimal process fidelity.

\begin{table*}[ht]
\caption{``Worst case'' estimate of process matrix elements for
the experimental data from \cite{Oka05} shown in table \ref{tbl2}.}
\label{tbl4}
\begin{center} \large
\begin{tabular}{l|cccc|c}
$\chi_{(f_z,f_x),(f_z,f_x)}$ & $f_x=0$ & $f_x=1$ & $f_x=2$ & $f_x=3$ & Sum
\\ \hline &&&& \\[-0.4cm]
$f_z=0$ & 0.720 & 0.034 & 0.071 & 0.028 & 0.853
\\
$f_z=1$ & 0.051 & 0 & 0 & 0 & 0.051
\\
$f_z=2$ & 0.052 & 0 & 0 & 0 & 0.052
\\
$f_z=3$ & 0.044 & 0 & 0 & 0 & 0.044
\\[0.1cm] \hline &&&& \\[-0.4cm] Sum
& 0.867 & 0.034 & 0.071 & 0.028 & 1
\end{tabular}
\normalsize
\end{center}
\end{table*}

Another operation of great interest is the generation of
entanglement from local $XZ$ inputs. The ideal operation
$\hat{U}_0$ converts these input states into the four
orthogonal Bell states, characterized by their $XX$, $YY$,
and $ZZ$ correlations. The fidelity $F_{\mbox{E1}}$ of
this operation is also given by a sum of four process matrix
elements, corresponding to the Bell state stabilizers
$II$, $XX$, $YY$, and $ZZ$,
\begin{equation}
\label{eq:fe1}
F_{\mbox{E1}}= F_{qp} + \chi_{(3,0)(3,0)} + \chi_{(0,3)(0,3)}
+ \chi_{(3,3)(3,3)}.
\end{equation}
The measurement results define a minimal fidelity for this
operation given by one minus the single qubit errors $f_z=1$,
$f_z=2$, $f_x=1$ and $f_x=2$,
\begin{equation}
F_{\mbox{E1}} \geq 1 - \eta_Z(1) - \eta_Z(2) - \eta_X(1) -\eta_X(2)
= 0.792.
\end{equation}
It is therefore possible to improve the estimate of entanglement
capability originally reported in \cite{Oka05} without adding any
new data.

It may also be interesting to apply the same analysis to
the reverse operation, that converts the Bell states into
local $XZ$ eigenstates. The fidelity $F_{xz}$ of this
disentangling operation is given by the process matrix
element sum
\begin{equation}
\label{eq:fxz}
F_{xz}= F_{qp} + \chi_{(2,0)(2,0)} + \chi_{(0,1)(0,1)}
+ \chi_{(2,1)(2,1)},
\end{equation}
and the minimal fidelity defined by the measurement results
is
\begin{equation}
F_{xz} \geq 1 - \eta_Z(1) - \eta_Z(3) - \eta_X(2) -\eta_X(3)
= 0.806.
\end{equation}
The minimal fidelity of disentangling the Bell states is thus
higher than the minimal fidelity of entanglement generation
by the time-reversed process.

Finally, it is also possible to consider the errors in entanglement
generation from $YY$ inputs. The outputs are then maximally entangled
states with the stabilizers $II$, $ZY$, $YX$, and $XZ$.
Therefore, the fidelity $F_{\mbox{E2}}$ is given by
\begin{equation}
\label{eq:fe2}
F_{\mbox{E2}}= F_{qp} + \chi_{(1,3)(1,3)} + \chi_{(3,2)(3,2)}
+ \chi_{(2,1)(2,1)}.
\end{equation}
In this case, all of the stabilizers represent errors that show
up in both the $Z$ and the $X$ measurements. Therefore, the
fidelity $F_{xz}$ can be as low as the minimal process fidelity,
\begin{equation}
F_{\mbox{E2}} \geq F_Z+F_X-1 = 0.720.
\end{equation}
Thus the measurement data provided by the complementary
$Z$ and $X$ operations provides only a very rough estimate
of the operation on $Y$ inputs.

Of course, the worst case noise estimate given in table \ref{tbl4}
is a rather extreme and unlikely interpretation of the measurement
data. It may therefore be interesting to compare it with a more
realistic estimate based on the assumption that the errors
in $Z$ and $X$ are uncorrelated \cite{Hof06b}.
For this purpose, it is useful to consider the definition of
the average fidelity of a quantum process,
\begin{equation}
F_{\mbox{av.}} = \int d\Phi
\langle \Phi \mid \hat{U}_0^\dagger
E(\mid \Phi \rangle \langle \Phi \mid) \hat{U}_0 \mid \Phi \rangle,
\end{equation}
where the integral over $\Phi$ represents a uniform average over
all possible states of the $d$-dimensional Hilbert space.
It has been shown \cite{Hor99} that this average fidelity is related to the
process fidelity by
\begin{equation}
\label{eq:FavFqp}
F_{\mbox{av.}} = \frac{F_{qp}\; d +1}{d+1}.
\end{equation}
This relation can be explained quite intuitively by using the error
analysis above. As eq.(\ref{eq:relate}) indicates, the fidelities
for one input basis of $d$ orthogonal states are given by a sum of
the process fidelity $F_{qp}$ and the diagonal process matrix elements
of $d-1$ errors, out of a total number of $d^2-1=(d+1)(d-1)$ possible
errors. This means that the probability that the operation on any
given input state $\mid \Phi \rangle$ is insensitive to any given
error is $(d-1)/(d^2-1)=1/(d+1)$. The average over all possible
input states $\mid \Phi \rangle$ makes this relation exact: the
contribution of $1/(d+1)$ to $F_{\mbox{av.}}$ in eq.(\ref{eq:FavFqp})
simply represents the probability that the errors in the quantum
operation result only in an unobservable phase change.

\begin{table*}[ht]
\caption{Statistical estimate of the process matrix elements for
the experimental data from \cite{Oka05} shown in table \ref{tbl2}.
The assumption of uncorrelated errors results in a uniform distribution
of errors over all possibilities.}
\label{tbl5}
\begin{center} \large
\begin{tabular}{l|cccc|c}
$\chi_{(f_z,f_x),(f_z,f_x)}$ & $f_x=0$ & $f_x=1$ & $f_x=2$ & $f_x=3$ & Sum
\\ \hline &&&& \\[-0.4cm]
$f_z=0$ & 0.825 & 0.0072 & 0.0150 & 0.0059 & 0.853
\\
$f_z=1$ & 0.0146 & 0.0093 & 0.0194 & 0.0077 & 0.051
\\
$f_z=2$ & 0.0149 & 0.0095 & 0.0198 & 0.0078 & 0.052
\\
$f_z=3$ & 0.0126 & 0.0080 & 0.0168 & 0.0066 & 0.044
\\[0.1cm] \hline &&&& \\[-0.4cm] Sum
& 0.867 & 0.034 & 0.071 & 0.028 & 1
\end{tabular}
\normalsize
\end{center}
\end{table*}

This interpretation of eq.(\ref{eq:FavFqp}) indicates that we can
consider the complementary fidelities $F_X$ and $F_Z$ as
representative fidelities contributing to the average
fidelity $F_{\mbox{av.}}$. If the complementary fidelities are not
too different from each other, the most realistic assumption
seems to be that the average fidelity $F_{\mbox{av.}}$ is close to the
average of the complementary fidelities. An estimate of the most
likely value of the process fidelity $F_{qp}$ can then be obtained
from
\begin{equation}
F_{qp}(\mbox{est.}) \approx \left(1+\frac{1}{d}\right)
\left(\frac{F_Z+F_X}{2}\right)-\frac{1}{d}.
\end{equation}
Since this estimate implicitly assumes a rather uniform distribution
of errors, it is natural to apply the same assumption to obtain
an estimate for the remaining diagonal elements of the process matrix.
This means that the joint probabilities of errors given by
$\chi_{(f_z,f_x)(f_z,f_x)}$ should be proportional to the product
of the error probabilities $\eta_Z(f_z)$ and $\eta_X(f_x)$ for
$f_z,f_x \neq 0$. With this assumption of uncorrelated errors,
the remaining process matrix elements read
\begin{eqnarray}
\lefteqn{\chi_{(f_z,f_x)(f_z,f_x)}(\mbox{est.}) \approx}
\nonumber \\ &&
\frac{d-1}{d}\left(\frac{1}{1-F_Z}+\frac{1}{1-F_X}\right)
\eta_Z(f_z)\eta_X(f_x)
\nonumber \\
\lefteqn{\chi_{(f_z,0)(f_z,0)}(\mbox{est.}) \approx}
\nonumber \\ &&
\left(
\frac{d+1}{2 d}-\frac{d-1}{2 d} \left(\frac{1-F_X}{1-F_Z}\right)
\right) \eta_Z(f_z)
\nonumber \\
\lefteqn{\chi_{(0,f_x)(0,f_x)}(\mbox{est.}) \approx}
\nonumber \\ &&
\left(
\frac{d+1}{2 d}-\frac{d-1}{2 d} \left(\frac{1-F_Z}{1-F_X}\right)
\right) \eta_X(f_x).
\end{eqnarray}
This estimate may be the most realistic description for quantum
processes with very similar complementary fidelities $F_Z$ and
$F_X$. If the values of $F_Z$ and $F_X$ are very different, it
is possible that the estimates for $\chi_{(0,f_x)(0,f_x)}$
or for $\chi_{(f_z,0)(f_z,0)}$ become negative, especially if
the higher fidelity is close to one. Although such errors can
be corrected by restricting all diagonal elements to positive
values, it would be more natural to use a different noise models
for such cases.

In the case of our controlled-NOT gate, the similar values of
$F_Z$ and $F_X$ indicate that the statistical noise model may
be appropriate. The corresponding distribution of process
matrix elements is shown in table \ref{tbl5}. All diagonal
elements representing errors now have very similar values,
ranging from a minimum of 0.59 \% for $(0,3)$ to a maximum of
1.98 \% for $(2,2)$. The estimated process
fidelity of $F_{qp}=0.825$ is only a little bit lower than the
maximal possible process fidelity of $\mbox{Min}\{F_Z,F_X\}=0.853$,
and much higher than the lower limit of $F_Z+F_Z-1=0.720$.
Statistical considerations thus indicate that the actual process
fidelity of the device is likely to significantly exceed the minimum
assumed in the ``worst case'' estimate shown in table \ref{tbl4}.

We can also derive estimates for other operations from the
statistical noise model, corresponding to their most likely
values. The results derived from eqs. (\ref{eq:fzx},
\ref{eq:fe1},\ref{eq:fxz},\ref{eq:fe2}) read
\begin{eqnarray}
F_{zx}\approx 0.874 && F_{\mbox{E1}}\approx 0.850
\nonumber \\
F_{xz}\approx 0.857 && F_{\mbox{E2}}\approx 0.859.
\end{eqnarray}
Naturally, all of these fidelities are now close
to the average fidelity of $F_{\mbox{av.}}=(F_Z+F_X)/2=0.86$.
Interestingly, the estimate for the fidelity $F_{\mbox{E2}}$
of entanglement generation from $YY$ inputs is now higher
than the estimate for the fidelity $F_{\mbox{E1}}$ of
entanglement generation from $XZ$ inputs. This change
illustrates the fundamental difference between minimal values
and likely values in the fidelity estimates. In fact, the
two results show that the available data allows a far more
precise estimate of $F_{\mbox{E1}}$ than of $F_{\mbox{E2}}$.

\section{Conclusions}

The errors observed in $N$-qubit operations can be
characterized in terms of the bit flip errors observed
in the complementary operations resulting in $Z$ and in
$X$ output states. It is therefore useful to characterize
a device by first measuring the $2 d = 2^{N+1}$ output
fidelities of these two complementary operations.
Since the only process able to perform both operations
with a fidelity of $1$ is the intended process $\hat{U}_0$,
it is possible to derive upper and lower bounds for the
process fidelity from these two measurements.
For high fidelity processes, this estimate is sufficient to
confirm the successful implementation of a multi qubit gate.

For noisy processes such as the quantum controlled-NOT
analyzed above, the details results for the fidelities and
errors allow estimates of the process matrix elements
corresponding to various noise models. Even though these
error models are not as precise as the estimates obtained
from full quantum tomography, it is remarkable that
such a detailed analysis is possible using only a small
fraction of the $d^4=16^N$ measurement probabilities required
for complete quantum process tomography. The evaluation of
complementary operations is therefore a particularly
efficient method for the characterization of multi-qubit
quantum devices.

\section*{Acknowledgements}
This work was partially supported by the CREST program of the
Japan Science and Technology Agency, JST, and the Grant-in-Aid
program of the Japanese Society for the Promotion of Science,
JSPS.

\end{document}